\documentclass[preprint]{sig-alternate}

\setcopyright{none}
\toappear{}

\pdfoutput=1
\newif\ifdraft\draftfalse
\ifdraft
  \usepackage{showlabels}
\else
  \usepackage{balance}
\fi

\usepackage[clock]{ifsym}

\usepackage[utf8]{inputenc}

\usepackage[T1]{fontenc}

\usepackage{setspace}

\usepackage{relsize}

\usepackage{lmodern}

\usepackage{stmaryrd}
\usepackage{textcomp}

\usepackage[normalem]{ulem}

\usepackage{cite}

\usepackage{booktabs}
\usepackage{dcolumn}
\newcolumntype{.}{D{.}{.}{-1}}
\newcolumntype{,}{D{,}{,}{-1}}
\usepackage[table]{xcolor}

\usepackage[flushleft,alwaysadjust]{paralist}

\usepackage{stfloats}

\usepackage{nonfloat}

\usepackage{amsmath}
\usepackage{amssymb}

\usepackage{mathtools}

\usepackage{mathdots}

\usepackage{colonequals}

\usepackage[font=bf]{caption}
\usepackage[labelformat=parens,font+=normalsize]{subcaption}
\usepackage{url}
\usepackage{graphicx}

\usepackage{tikz}
\usetikzlibrary{arrows.meta}
\usetikzlibrary{patterns}
\usetikzlibrary{shapes.symbols}
\usetikzlibrary{shapes.misc}
\usetikzlibrary{calc}
\pgfdeclarelayer{background}
\pgfdeclarelayer{foreground}
\pgfsetlayers{background,main,foreground}
\usepackage{pgfplots}
\usepackage{pgfmath}

\usepackage{bigstrut}
\usepackage{array}
\newcommand{\col}[1]{\ensuremath{\sql{#1}}}
\newcolumntype{H}{>{\columncolor{black}\color{white}}c}
\newcolumntype{P}{>{\columncolor{white}\color{white}}c}
\newcommand{\colhd}[1]{\multicolumn{1}{H}{\strut\col{#1}}}

\newcommand{\keyhd}[1]{\multicolumn{1}{H}{\strut\underline{\smash{\col{#1}}}}}
\newenvironment{littbl}{%
  \renewcommand{\arraycolsep}{1pt}%
  \renewcommand{\tabcolsep}{1pt}\ttfamily\ignorespaces}{\par\ignorespacesafterend}
\newcommand{\tabname}[2]{%
  \multicolumn{#1}{@{}l}{%
    \begin{tikzpicture}[anchor=center,baseline]
      \node [inner sep=1pt] (T) {\phantom{#2}};
      \draw [fill=black]
            (T.south west) {[rounded corners=2pt] -- (T.north west)  --
            (T.north east)} -- (T.south east) -- cycle;
      \node [inner sep=1pt,text=white] {#2};
    \end{tikzpicture}}}

\usepackage{fancyvrb}
\usepackage{listings}
\makeatletter
\def\lst@visiblespace{\textnormal{\textvisiblespace\,}}
\makeatother
\let\origthelstnumber\thelstnumber
\makeatletter
\newcommand{\suppressnumber}{%
  \lst@AddToHook{OnNewLine}{%
    \let\thelstnumber\relax%
     \advance\c@lstnumber-\@ne\relax}}
\newcommand{\reactivatenumber}{%
  \lst@AddToHook{OnNewLine}{%
   \let\thelstnumber\origthelstnumber%
   \advance\c@lstnumber\@ne\relax}}
\newcommand{\forcenumber}[1]{%
  \lst@AddToHook{OnNewLine}{%
   \let\thelstnumber\origthelstnumber%
   \setcounter{lstnumber}{\numexpr#1-1\relax}}}
\makeatother
\makeatletter
\lst@Key{countblanklines}{true}[t]%
    {\lstKV@SetIf{#1}\lst@ifcountblanklines}
\lst@AddToHook{OnEmptyLine}{%
    \lst@ifnumberblanklines\else%
       \lst@ifcountblanklines\else%
         \advance\c@lstnumber-\@ne\relax%
       \fi%
    \fi}
\makeatother
\lstdefinelanguage{sql}{
  keywordstyle={[1]\bfseries},
  keywordstyle={[2]\ttfamily},
  countblanklines=false,
  numberblanklines=false,
  morestring=[d]{'},
  morecomment=[l]{--},
  morekeywords={CREATE,FUNCTION,RETURNS,AS,LANGUAGE,SQL,PLPGSQL,STABLE,IMMUTABLE,
    DECLARE,BEGIN,END,FOR,LOOP,IN,IF,THEN,ELSE,RETURN,
    WITH,RECURSIVE,UNION,ALL,
    SELECT,FROM,WHERE,GROUP,ORDER,BY,SUM,AND,OR,COALESCE,BETWEEN,
    WINDOW,OVER,ROWS,UNBOUNDED,PRECEDING,EXCLUDE,CURRENT,ROW,
    LATERAL,CASE,WHEN},
  morekeywords=[2]{sign}
}
\lstdefinelanguage{ssaanf}{
  keywordstyle={[1]\scshape},
  keywordstyle={[2]\ttfamily},
  countblanklines=false,
  numberblanklines=false,
  morestring=[d]{'},
  morecomment=[l]{--},
  morekeywords={function,if,then,else,return,goto,
    letrec,let,rec,in,
    SELECT,OR},
  morekeywords=[2]{sign}
}

\usepackage{fancybox}
\makeatletter
\newenvironment{lstbox}{%
  \begin{Sbox}\ignorespaces}{%
  \end{Sbox}\parbox[t]{\wd\@Sbox}{\TheSbox}\ignorespacesafterend}
\makeatother

\usepackage[activate=false]{microtype}

\usepackage[capitalise]{cleveref}

\crefname{section}{Section}{Sections}
\crefname{subsection}{Section}{Sections}
\crefname{subsubsection}{Section}{Sections}
\crefname{appendix}{Appendix}{Appendices}
\crefname{equation}{Equation}{Equations}
\crefname{figure}{Figure}{Figures}
\crefname{table}{Table}{Tables}
\crefname{subfigure}{Figure}{Figures}
\crefname{subtable}{Table}{Tables}
\crefformat{item}{\itshape #1.}

\newlength{\x}
\newlength{\y}
\newlength{\xx}

\newcommand{\SQL}    {S\kern-0.06emQ\kern-0.06emL}
\newcommand{\PLSQL}  {P\kern-0.09emL\kern-0.07em/\kern-0.07em\SQL}
\newcommand{\PLpgSQL}{P\kern-0.09emL\kern-0.07em/\kern-0.07emp\kern-0.07emg\kern-0.07em\SQL}
\newcommand{\Pg}     {Postgre\SQL}
\newcommand{\Oracle} {Oracle}               %% DeWitt?!

\newcommand{\sql}[1]{\textnormal{\ttfamily\textls[-80]{#1}}\/}

\newcommand{\var}[1]{\mathit{#1}}

\pgfdeclarepatternformonly{stripes}{\pgfpoint{0cm}{0cm}}{\pgfpoint{1.2cm}{1.2cm}}{\pgfpoint{1.2cm}{1.2cm}}
{
  \foreach \i in {0.0, 0.2, 0.4, 0.6, 0.8, 1.0, 1.2}
  {
     \pgfpathmoveto{\pgfpoint{\i cm}{0cm}}
     \pgfpathlineto{\pgfpoint{1.2cm}{1.2cm - \i cm}}
     \pgfpathlineto{\pgfpoint{1.2cm}{1.2cm - \i cm + 0.1cm}}
     \pgfpathlineto{\pgfpoint{\i cm - 0.1cm}{0cm}}
     \pgfpathclose%
     \pgfusepath{fill}
     \pgfpathmoveto{\pgfpoint{0cm}{\i cm}}
     \pgfpathlineto{\pgfpoint{1.2cm - \i cm}{1.2cm}}
     \pgfpathlineto{\pgfpoint{1.2cm - \i cm - 0.1cm}{1.2cm}}
     \pgfpathlineto{\pgfpoint{0cm}{\i cm + 0.1cm}}
     \pgfpathclose%
     \pgfusepath{fill}
   }
}

\DeclareFontFamily{OML}{cmtex}{}
\DeclareFontShape{OML}{cmtex}{m}{n}{%
  <8> <9> <10> gen * cmtex
  <5-> fixed * cmtex8 }{}
\DeclareSymbolFont{ops}{OML}{cmtex}{m}{n}
\DeclareMathSymbol{\ttleftarrow} {\mathord}{ops}{24}
\DeclareMathSymbol{\ttrightarrow}{\mathord}{ops}{25}
\DeclareMathSymbol{\ttuparrow}   {\mathord}{ops}{11}
\DeclareMathSymbol{\ttdownarrow} {\mathord}{ops}{1}

\newcommand{\ssa}[1]{\ensuremath{\mskip1mu_{\sql{#1}}}}
\newcommand{\SSA}[1]{\lstinline[language=ssaanf]{#1}}
\newcommand{\ANF}[1]{\textnormal{\lstinline[language=ssaanf]{#1}}}
\newcommand{\assign}{\ttleftarrow}
\newcommand{\free}[1]{\uline{#1}}
\newcommand{\wrapped}{\smash{\ensuremath{^{\texttt{*}}}}}

\newcommand{\frag}[1]{%
  \begin{tikzpicture}[baseline=(id.base)]
    \node[outer sep=0mm,inner sep=1pt,minimum width=2.5mm,minimum height=2.5mm,
          text=white,draw=white,line width=0.5pt,fill=black,rounded corners=0pt,
          font=\small\ttfamily] at (0,0) (id) {#1};
   \end{tikzpicture}}

\newcommand{\listof}[1]{\smash[t]{\ensuremath{\overline{#1}}}}

\newcounter{listingsline}

\newcommand{\gutterlineskip}{%
  \ignorespaces\rlap{\hskip-6.6pt%
    \begin{tikzpicture}[overlay,y=-\baselineskip]
      \draw[densely dotted] (0,0.5)--(0,0);
    \end{tikzpicture}}\ignorespacesafterend}

\tikzstyle{cp}=[draw,-{Latex[round,length=2pt,width=2pt]},
                shorten <=-0.2pt,shorten >=-0.2pt,line width=0.5pt,
                rounded corners=1pt]
\colorlet{cp}{black!30}
\newcommand{\syn}[2][-0.5ex]{%
  \begin{tikzpicture}[baseline=#1]
    \node[inner sep=0mm,text=white,font=\scriptsize\ttfamily] at (0,0) (cp) {#2};
    \begin{pgfonlayer}{background}
      \draw[white,fill=black!90,line width=0.3pt] (cp.center) circle [radius=0.85ex];
    \end{pgfonlayer}
  \end{tikzpicture}}

\newcommand{\ctxtsw}{\textcolor{black!60}{\hskip1pt\textbar\hskip1pt}}
\newcommand{\Qtof}{\ensuremath{\sql{Q}\ttrightarrow\sql{f}}}
\newcommand{\ftoQ}[1][f]{\ensuremath{\sql{#1}\ttrightarrow\sql{Q}_i}}

\newcommand{\step}[1]{%
  \begin{tikzpicture}[inner sep=1pt,baseline=(b.base)]
    \node[rounded rectangle,minimum width=9.5mm,fill=black!30,text=black,text depth=0pt,font=\bfseries] (b) {#1};
  \end{tikzpicture}}

\newcommand{\compile}[1]{\llbracket #1\rrbracket}

\newlength{\G}

\tikzstyle{envelope}=[fill=black!20,rounded corners=1pt]

\begin{document}

\title{Compiling \PLSQL{} Away \\[-4mm]}

\numberofauthors{1}
\author{
  \alignauthor
  Christian Duta
  \qquad
  Denis Hirn
  \qquad
  Torsten Grust
  \\[1ex]
  \affaddr{Universit\"at T\"ubingen} \\
  \affaddr{T\"ubingen, Germany}
  \\[1ex]
  \email{[ christian.duta, denis.hirn, torsten.grust ]@uni-tuebingen.de}
}

\maketitle

\begin{abstract}
  ``\PLSQL{} functions are slow,'' is common developer wisdom that
  derives from the tension between set-oriented \SQL{} evaluation and
  statement-by-statement \PLSQL{} interpretation.  We pursue the radical
  approach of compiling \PLSQL{} away, turning interpreted functions
  into regular subqueries that can then be efficiently evaluated
  together with their embracing \SQL{} query, avoiding any
  \PLSQL{}\,$\leftrightarrow$\,\SQL{} context switches.  Input \PLSQL{}
  functions may exhibit arbitrary control flow.  Iteration, in
  particular, is compiled into \SQL{}-level recursion.  RDBMSs across
  the board reward this compilation effort with significant run time
  savings that render established developer lore questionable.
\end{abstract}
% Translate iteration into tail recursion.  Common wisdom points into the
% opposite direction.

\section{Now Is Not a Good Time \\ to Interrupt Me}
\label{sec:introduction}

\noindent
Frequent changes\ctxtsw{}The required\ctxtsw{}of context\ctxtsw{}context
switching effort\ctxtsw{}can turn\ctxtsw{}may even outweigh\ctxtsw{}otherwise
tractable tasks\ctxtsw{}the cost\ctxtsw{}into real challenges.\ctxtsw{}of the
tasks themselves.

\smallskip\noindent
If you have found those two sentences hard to comprehend, you were
struggling with the \emph{context switches}---occurring at
every\,\ctxtsw{}\,bar---needed to process a piece of one sentence before
immediately turning focus back to the other.

\smallskip\noindent
\SQL{} evaluation in relational DBMSs can face such frequent context
switches, in particular if bits of the query logic are implemented using
\PLSQL,\footnote{We refer to the language as \emph{\PLSQL{}} as coined by Oracle.
Our discussion extends to its variants known as \emph{\PLpgSQL} in \Pg{}
or \emph{T-\SQL} in Microsoft \SQL{} Server.}
\emph{the} in-database scripting language.  Whenever a \SQL{} query~\sql{Q} invokes
a \PLSQL{} function, say~\sql{f},
\begin{compactitem}
\item the DBMS switches from set-oriented plan evaluation to statement-by-statement
  \PLSQL{} interpretation mode (referred to as switch~\Qtof{} in the sequel).
\item Execution of \sql{f}'s statements then
  switches query evaluation back to plan mode---possibly multiple
  times---to evaluate the \SQL{} queries~$\sql{Q}_i$ embedded in~\sql{f} (switch~\ftoQ).
\end{compactitem}
Context switches will be abundant.  If \sql{f}'s~call site is
located inside a~\sql{SELECT}-\sql{FROM}-\sql{WHERE} block of~\sql{Q},
each row processed by the block will invoke~\sql{f}.  Likewise, if \sql{f} embeds
multiple queries or employs iteration, \emph{e.g.}, in terms
of~\sql{FOR} or~\sql{WHILE} loops, we observe repeated plan evaluation
for the~$\sql{Q}_i$.

\smallskip\noindent
Unfortunately, both kinds of context switches are costly.
Each switch~\Qtof{} incurs overhead for \PLSQL{} interpreter invocation
or resumption.  A switch~\ftoQ{} leads to overhead due to
\begin{inparaenum}[(1)]
\item plan generation and caching on the first evaluation of~$\sql{Q}_i$ or
\item plan cache lookup, plan instantiation, and plan teardown for each subsequent evaluation of~$\sql{Q}_i$.
\end{inparaenum}
Iteration in both, \sql{Q} and~\sql{f}, multiplies the toll.

\begin{figure}[t]
  \begin{narrow}{-2mm}{-2mm}
  \centering\small
  \begin{subfigure}[b]{0.32\linewidth}
    \centering
    \begin{tikzpicture}[x=4mm,y=-4mm]
      \begin{scope}[xshift=-2mm,yshift=2mm]
        %% cell grid
        \draw[xstep=1,ystep=-1] (0,0) grid (5,5);
        %% walls
        \begin{scope}[pattern=stripes]
          \pattern (0,0) rectangle (2,1);
          \pattern (3,1) rectangle (4,2);
          \pattern (2,3) rectangle (3,4);
        \end{scope}
      \end{scope}
      %% coordinates
      \begin{scope}[black!70,font=\scriptsize]
        \node foreach \y in {0,...,4} at (-1,\y) {\y};
        \node at (-1,5) {$y$};
        \node foreach \x in {0,...,4} at (\x,-1) {\x};
        \node at (5,-1) {$x$};
      \end{scope}
      %% rewards
      \begin{scope}[text depth=0pt]
        \node at (2,0) {$-2$};
        \node at (3,0) {$0$};
        \node at (4,0) {$-1$};
        \node at (0,1) {$-2$};
        \node at (1,1) {$0$};
        \node at (2,1) {$-1$};
        \node at (4,1) {$1$};
        \node at (0,2) {$1$};
        \node at (1,2) {$1$};
        \node at (2,2) {$-1$};
        \node at (3,2) {$-1$};
        \node at (4,2) {$0$};
        \node at (0,3) {$-2$};
        \node at (1,3) {$1$};
        \node at (3,3) {$0$};
        \node at (4,3) {$-1$};
        \node at (0,4) {$-1$};
        \node at (1,4) {$0$};
        \node at (2,4) {$-1$};
        \node at (3,4) {$-2$};
        \node at (4,4) {$-2$};
      \end{scope}
    \end{tikzpicture}
    \caption{Cell rewards.}
    \label{fig:cells}
  \end{subfigure}%
  \hfill\vrule\hfill
  \begin{subfigure}[b]{0.32\linewidth}
    \centering
    \begin{tikzpicture}[x=4mm,y=-4mm]
      \begin{scope}[xshift=-2mm,yshift=2mm]
        %% cell grid
        \draw[xstep=1,ystep=-1] (0,0) grid (5,5);
        %% walls
        \begin{scope}[pattern=stripes]
          \pattern (0,0) rectangle (2,1);
          \pattern (3,1) rectangle (4,2);
          \pattern (2,3) rectangle (3,4);
        \end{scope}
      \end{scope}
      %% coordinates
      \begin{scope}[black!70,font=\scriptsize]
        \node foreach \y in {0,...,4} at (-1,\y) {\y};
        \node at (-1,5) {$y$};
        \node foreach \x in {0,...,4} at (\x,-1) {\x};
        \node at (5,-1) {$x$};
      \end{scope}
      %% actions
      \begin{scope}[text depth=0pt,anchor=mid]
        \node at (2,0) {$\shortdownarrow$};
        \node at (3,0) {$\shortdownarrow$};
        \node at (4,0) {$\shortdownarrow$};
        \node at (0,1) {$\shortdownarrow$};
        \node at (1,1) {$\shortrightarrow$};
        \node at (2,1) {$\shortdownarrow$};
        \node at (4,1) {$\shortdownarrow$};
        \node at (0,2) {$\shortdownarrow$};
        \node at (1,2) {$\shortdownarrow$};
        \node at (2,2) {$\shortleftarrow$};
        \node at (3,2) {$\shortrightarrow$};
        \node at (4,2) {$\shortuparrow$};
        \node at (0,3) {$\shortuparrow$};
        \node at (1,3) {$\shortrightarrow$};
        \node at (3,3) {$\shortleftarrow$};
        \node at (4,3) {$\shortuparrow$};
        \node at (0,4) {$\shortrightarrow$};
        \node at (1,4) {$\shortuparrow$};
        \node at (2,4) {$\shortleftarrow$};
        \node at (3,4) {$\shortuparrow$};
        \node at (4,4) {$\shortuparrow$};
      \end{scope}
    \end{tikzpicture}
    \caption{Markov policy.}
    \label{fig:policy}
  \end{subfigure}%
  \hfill\vrule\hfill
  \begin{subfigure}[b]{0.34\linewidth}
    \centering
    \begin{tikzpicture}[x=10mm,y=-10mm]
      %% cells
      \begin{scope}[minimum width=6mm,minimum height=6mm,inner sep=0pt,font=\scriptsize]
        \node[draw,pattern=stripes,rectangle] at (0,-1) (wall) {};
        \node[draw,rectangle,align=center]    at (0, 0) (c32)  {(3,2)};
        \node[draw,rectangle]                 at (0, 1) (c33)  {(3,3)};
        \node[draw,rectangle]                 at (1, 0) (c42)  {(4,2)};
      \end{scope}
      \node[anchor=west,right=3mm] at (wall) {wall};
      %% actions
      \begin{scope}[line cap=round,densely dotted,shorten <=0.5mm,shorten >=0.5mm,font=\tiny,text=black!70]
        \draw[->,line width=0.8pt]    (c32) -- (c42)                               node[pos=0.5,below=1pt] {0.8};
        \draw[->]                     (c32) -- (c33)                               node[pos=0.5,left]      {0.1};
        \draw[->,rounded corners=2pt] (c32) -- ++(-0.1,-0.65) -- +(0.2,0) -- (c32) node[pos=0.5,left=3pt]  {0.1};
      \end{scope}
      %% robot
      \node at (-0.7, 0.15) {\includegraphics[width=5mm]{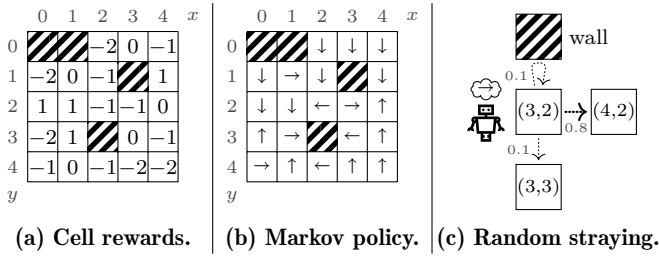}};
      \node[draw,cloud,cloud puffs=8,inner sep=-1pt,minimum width=4mm] at (-0.7,-0.3) {$\shortrightarrow$};
    \end{tikzpicture}
    \caption{Random straying.}
    \label{fig:actions}
  \end{subfigure}
  \end{narrow}
  \caption{Controlling an unreliable robot to collect rewards.}
  \label{fig:cells-policy-actions}
\end{figure}

\begin{figure}
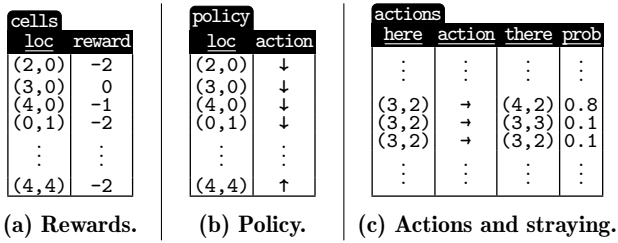

  \centering\small
  \begin{subfigure}[b]{0.26\linewidth}
    \centering
    \begin{littbl}
       \begin{tabular}{@{}|c|c|@{}}
         \tabname{2}{\sql{cells}} \\
         \keyhd{loc} & \colhd{reward} \\
         \strut
         (2,0) & -2  \\
         (3,0) &  ~0 \\
         (4,0) & -1  \\
         (0,1) & -2  \\
         \raisebox{3pt}{$\vdots$} & \raisebox{3pt}{$\vdots$} \\
         (4,4) & -2  \\
         \hline
       \end{tabular}
    \end{littbl}
    \caption{Rewards.}
    \label{fig:cells-table}
  \end{subfigure}%
  \hfill\vrule\hfill
  \begin{subfigure}[b]{0.26\linewidth}
    \centering
    \begin{littbl}
       \begin{tabular}{@{}|c|c|@{}}
         \tabname{2}{\sql{policy}} \\
         \keyhd{loc} & \colhd{action} \\
         \strut
         (2,0) & $\ttdownarrow$ \\
         (3,0) & $\ttdownarrow$ \\
         (4,0) & $\ttdownarrow$ \\
         (0,1) & $\ttdownarrow$ \\
         \raisebox{3pt}{$\vdots$} & \raisebox{3pt}{$\vdots$} \\
         (4,4) & $\ttuparrow$   \\
         \hline
       \end{tabular}
    \end{littbl}
    \caption{Policy.}
    \label{fig:policy-table}
  \end{subfigure}%
  \hfill\vrule\hfill
  \begin{subfigure}[b]{0.425\linewidth}
    \centering
    \begin{littbl}
       \begin{tabular}{@{}|c|c|c|c|@{}}
          \tabname{2}{\sql{actions}} \\
          \keyhd{here} & \keyhd{action} & \keyhd{there} & \keyhd{prob} \\
          \strut
          \raisebox{3pt}{$\vdots$} & \raisebox{3pt}{$\vdots$} & \raisebox{3pt}{$\vdots$} & \raisebox{3pt}{$\vdots$} \\
          (3,2) & $\ttrightarrow$ & (4,2) & 0.8 \\
          (3,2) & $\ttrightarrow$ & (3,3) & 0.1 \\
          (3,2) & $\ttrightarrow$ & (3,2) & 0.1 \\
          \raisebox{3pt}{$\vdots$} & \raisebox{3pt}{$\vdots$} & \raisebox{3pt}{$\vdots$} & \raisebox{3pt}{$\vdots$} \\
          \hline
       \end{tabular}
    \end{littbl}
    \caption{Actions and straying.}
    \label{fig:actions-table}
  \end{subfigure}
  \caption{A tabular encoding of the robot control scenario.}
  \label{fig:cells-actions-policy-tables}
\end{figure}

\smallskip\noindent
Let us make the conundrum concrete with \PLpgSQL{} function~\sql{walk()}
of~\cref{fig:walk-plsql}. The function simulates the walk of a
robot~\smash{\raisebox{-1pt}{\includegraphics[width=2ex]{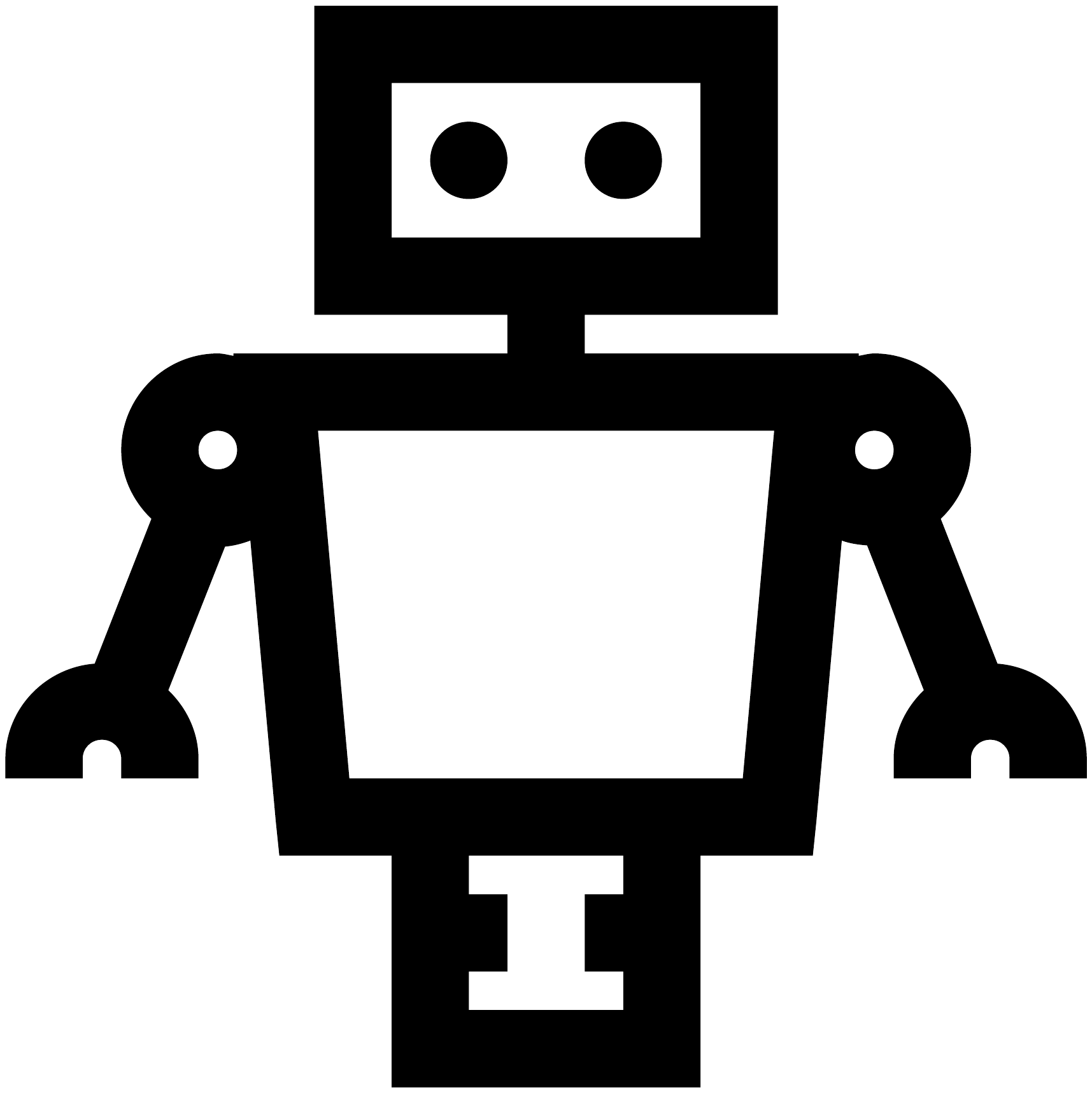}}}
on a grid whose cells hold rewards
(see~\cref{fig:cells,fig:cells-table}).  On cell~$(x,y)$ the robot
follows a prescribed policy (\emph{e.g.}, move down~$\ttdownarrow$ if on
cell~$(3,0)$, see~\cref{fig:policy,fig:policy-table}). This policy has
been precomputed by a Markov decision process which takes into account that
the robot may stray from its prescribed path: a planned move right from
$(3,2)$ will reach $(4,2)$ with probability $80\%$ but may actually
end up in $(3,3)$ or $(3,2)$, each with probability $10\%$
(see~\cref{fig:actions,fig:actions-table}). A
call~\sql{walk($o$,$w$,$l$,$s$)} starts the robot in origin cell~$o$ and
performs a maximum of~$s$ steps; \sql{walk} returns early if the
accumulated reward exceeds~$w$ or falls below~$l$.

\newsavebox{\profilebar}
\sbox{\profilebar}{%
  \begin{tikzpicture}[x=1ex,y=-1ex]
    \fill[black!40] (0,0.3) rectangle +(3,0.7);
    \draw[black,line cap=round,line width=0.6pt] (1.2,-0.1) -- +(0,1.5);
    \fill[black] (0,0.3) rectangle +(1.2,0.7);
  \end{tikzpicture}}
\smallskip\noindent
Each execution of \PLSQL{}~function \sql{walk} leads to the iterated
evaluation of the embedded \SQL{} queries~$\sql{Q}\ssa{1\dots 3}$.  The
run time profile on the rightmost edge of~\cref{fig:walk-plsql}
identifies these embedded queries to use the lion share of execution
time (\emph{e.g.}, $\sql{Q}\ssa{2}$ accounts for~$54.02\%$ of \sql{walk}'s
overall run time). While we expect such embedded queries to dominate
over the evaluation of simpler expressions and statements, the profile
also shows that a significant portion of the evaluation time for the
$\sql{Q}_i$ stems from \ftoQ[walk]~context switch overhead
(see the black section~\usebox{\profilebar}
of the profile bars).  For \Pg{}, this cost is to be attributed to the
engine's \sql{Exe\-cutor\-Start} and~\sql{Exe\-cutor\-End} functions. These
prepare the $\sql{Q}_i$'s plans (\emph{i.e.}, copy the cached plan into
a runtime data structure and instantiate the query's placeholders) and
free temporary memory contexts, respectively.  The \sql{FOR}~loop
iteration in~\sql{walk} multiplies this effort. The bottom line shows
that \Pg{} invests more than~$35\%$ of its time in \ftoQ[walk] overhead
during each invocation of~\sql{walk}. \cref{sec:experiments} shows
similar or worse bad news for more \PLpgSQL{} functions.

\smallskip\noindent
\textbf{\itshape Froid.} \PLSQL{} has long been identified as a culprit for disappointing
database application performance and it is common developer wisdom to
``avoid
\PLSQL{} functions altogether if possible''~\cite{froid}.  The situation is dire
and has led to recent drastic
efforts---coined~\emph{Froid}~\cite{froid}---by the Microsoft \SQL{}
Server team: if function~\sql{f} \textbf{is simple enough}, compile
its statements into a regular
\SQL{} subquery~$\sql{Q}\ssa{f}$ that can be inlined into the containing
\SQL{} query~\sql{Q}. Queries~\sql{Q} and $\sql{Q}\ssa{f}$ may then be
planned once and executed together in set-oriented fashion, avoiding
any \Qtof{} or
\ftoQ{} overheads.  \SQL{} Server with \emph{Froid} indeed enjoys noticeable
performance improvements and has been recognized as a major step forward
by both, the developer as well as the database research
communities~\cite{brent-ozar}.

%% sketches of MDP

%% relational representation of MDP

%% Original walk() PL/SQL function
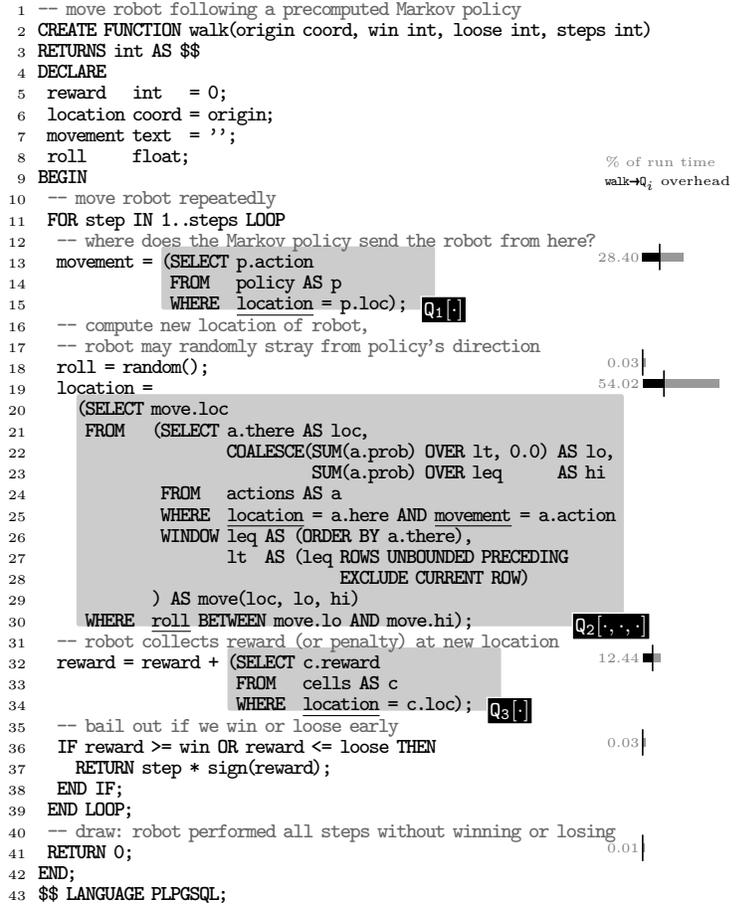
\begin{figure}[t]
  \centering\small
  %% measure the width of xx in a listing
  \settowidth{\xx}{\lstinline[columns=fixed]{xx}}\setlength{\x}{0.5\xx}
  \setlength{\y}{\baselineskip}\addtolength{\y}{-1.01pt}
  \begin{tikzpicture}[x=\x,y=-\y,inner sep=0mm]
    %% check width/height of character boxes
    %% \draw[help lines,xstep=1,ystep=-1] (0,0) grid (67,43);
    \node[anchor=north west] at (0,0) {%
      \begin{lstlisting}[language=sql,escapechar=\%]
-- move robot following a precomputed Markov policy
CREATE FUNCTION walk(origin coord, win int, loose int, steps int)
RETURNS int AS $$
DECLARE
 reward   int   = 0;
 location coord = origin;
 movement text  = '';
 roll     float;
BEGIN
 -- move robot repeatedly
 FOR step IN 1..steps LOOP
  -- where does the Markov policy send the robot from here?
  movement = (SELECT p.action
              FROM   policy AS p
              WHERE  %\free{\sql{location}}% = p.loc);
  -- compute new location of robot,
  -- robot may randomly stray from policy's direction
  roll = random();
  location =
    (SELECT move.loc
     FROM   (SELECT a.there AS loc,
                    COALESCE(SUM(a.prob) OVER lt, 0.0) AS lo,
                             SUM(a.prob) OVER leq      AS hi
             FROM   actions AS a
             WHERE  %\free{\sql{location}}% = a.here AND %\free{\sql{movement}}% = a.action
             WINDOW leq AS (ORDER BY a.there),
                    lt  AS (leq ROWS UNBOUNDED PRECEDING
                                EXCLUDE CURRENT ROW)
            ) AS move(loc, lo, hi)
     WHERE  %\free{\sql{roll}}% BETWEEN move.lo AND move.hi);
  -- robot collects reward (or penalty) at new location
  reward = reward + (SELECT c.reward
                     FROM   cells AS c
                     WHERE  %\free{\sql{location}}% = c.loc);
  -- bail out if we win or loose early
  IF reward >= win OR reward <= loose THEN
    RETURN step * sign(reward);
  END IF;
 END LOOP;
 -- draw: robot performed all steps without winning or losing
 RETURN 0;
END;
$$ LANGUAGE PLPGSQL;
\end{lstlisting}};
    \begin{scope}[rounded corners=1pt]
      \begin{pgfonlayer}{background}
        %% embedded SFW blocks
        \fill[black!20] (13,12) rectangle (42,15) node (Q1) {};
        \fill[black!20] ( 4,19) rectangle (62,30) node (Q2) {};
        \fill[black!20] (20,31) rectangle (49,34) node (Q3) {};
        \node[right=-2mm] at (Q1) {\frag{$\mathtt{Q_1}[\cdot]$}};
        \node[right=-7mm] at (Q2) {\frag{$\mathtt{Q_2}[\cdot,\cdot,\cdot]$}};
        \node[right=-2mm] at (Q3) {\frag{$\mathtt{Q_3}[\cdot]$}};
      \end{pgfonlayer}
    \end{scope}
    %% line-based profiling
    \begin{scope}[overlay,black!50]
      \node[anchor=west,font=\tiny]       at (60,8)  {\% of run time};
      \node[anchor=west,black,font=\tiny] at (60,9)  {\sql{walk}\raisebox{-1pt}{$\ttrightarrow$}$\sql{Q}_i$ overhead};
      \foreach \line/\perc/\disp/\overhead in {
        12/28.4 /28.40/12,
        17/2    / 0.03/ 0,
        18/54.02/54.02/15,
        31/12.44/12.44/ 7,
        35/2    / 0.03/ 0,
        40/1    / 0.01/ 0
      } {
        %% bar
        \pgfmathsetmacro\bar{0.15*\perc}
        %\draw (64,\line)+(0,0.1) -- ++(0,0.9);
        \fill[black!40] (64,\line)+(0,0.3) rectangle ++(\bar,0.7);
        %% measurement
        \node[anchor=east,xshift=-0.3\x,yshift=-0.5\y,font=\tiny] at (64,\line) {\disp};
        %% overhead (if any)
        \pgfmathsetmacro\bar{0.15*\overhead}
        \draw[black,line cap=round,line width=0.6pt] (64,\line)++(\bar,-0.1) -- ++(0,1.2);
        \fill[black] (64,\line)+(0,0.3) rectangle ++(\bar,0.7);
      };
    \end{scope}
  \end{tikzpicture}
  \caption{Original \PLpgSQL{} function~\sql{walk}. Black
    sections of the profile bars \usebox{\profilebar} quantify
    \ftoQ{} context switch overhead.}
  \label{fig:walk-plsql}
\end{figure}

\smallskip\noindent
In a nutshell, \emph{Froid} transforms sequences of \PLSQL{} assignment
statements into subqueries that are chained together with \SQL{}
Server's \sql{OUTER APPLY}~\cite{froid,outer-apply}.  The technique is
elegant and simple but comes with severe restrictions: foremost, the
just mentioned chaining will only work for functions~\sql{f} that
exhibit loop-less control flow.  This rules out \PLSQL{} functions
like~\sql{walk} that build on \sql{WHILE} or~\sql{FOR} iteration,
arguably core constructs in any imperative programming language.

\smallskip\noindent
\textbf{Compile \PLSQL{} away.}
We, too, subscribe to the drastic approach of~\emph{Froid}. However, we
also believe that efforts that aim to host complex computation inside
the DBMS and thus close to the data, need to support expressive
programming language dialects.  Control flow restrictions will be an
immediate show stopper for the majority of interesting computational
workloads.
The present research thus sets out
\begin{compactenum}[1.]
\item to completely compile \PLSQL{} functions~\sql{f} away, transforming them into
  regular \SQL{} queries~$\sql{Q}\ssa{f}$.  The \PLSQL{} functions may feature iteration---in fact
  any control flow is acceptable.  If~\sql{f} indeed contained iteration, $\sql{Q}\ssa{f}$
  will employ a recursive common table expression (CTE, \sql{WITH~RECURSIVE}) to
  express this in pure \SQL{}. No changes to the underlying DBMS are required (although
  modest local changes can provide another boost, see~\cref{sec:experiments}).
\item We study and quantify the run time impact of this compilation approach
  and the benefit of getting rid of \Qtof{} and \ftoQ{} context switches, in particular.
\item As a by-product, the approach enables in-database programming
  support for DBMSs like \SQL{}ite3 that previously lacked any \PLSQL{} support at all.
\end{compactenum}
\cref{sec:compiling-plsql-away}, the core of this paper, elaborates on
the compilation technique that turns iterative \PLSQL{} into recursive
\SQL{}. We hope to show that the transformation is systematic and practical.
Along the way, we point out several opportunities to make the approach even
more efficient. \cref{sec:experiments} reports on experimental observations
we made once we compiled \PLSQL{} away.

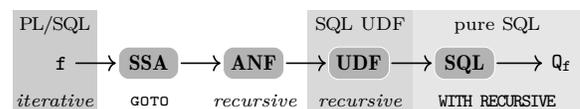
\begin{figure}[b]
  \centering\small
  \begin{tikzpicture}[x=1.4mm,y=-1mm,outer sep=1pt]
    %% forms
    \begin{scope}[anchor=east]
      \node at (0,0) (f) {\sql{f}};
      \begin{scope}[text=black,text depth=1pt,minimum width=8mm,rounded corners=4pt]
        \tikzset{every node/.style={draw=white,line width=0.3pt,fill=black!30}}
        \node at (10,0) (SSA) {\textbf{SSA}};
        \node at (20,0) (ANF) {\textbf{ANF}};
        \node at (30,0) (UDF) {\textbf{UDF}};
        \node at (40,0) (SQL) {\textbf{SQL}};
      \end{scope}
      \node at (48,0) (Qf) {\sql{Q}\ssa{f}};
    \end{scope}
    %% data flow
    \begin{scope}[line width=0.6pt,line cap=round]
      \draw[->] (f)   -- (SSA);
      \draw[->] (SSA) -- (ANF);
      \draw[->] (ANF) -- (UDF);
      \draw[->] (UDF) -- (SQL);
      \draw[->] (SQL) -- (Qf);
    \end{scope}
    %% PL regions
    \begin{pgfonlayer}{background}
      \fill[black!20] (-6,-7) rectangle ( 2,7);
      \fill[black!15] (22,-7) rectangle (32,7);
      \fill[black!10] (32,-7) rectangle (48,7);
      \begin{scope}[densely dashed,font=\scriptsize]
        \node at (-2,-5) {\PLSQL};
        \node at (27,-5) {\SQL{} UDF};
        \node at (40,-5) {pure \SQL};
        %% iterative/recursive?
        \node at (-2,5) {\emph{iterative}};
        \node at ( 7,5) {\SSA{goto}};
        \node[align=center] at (17,5) {\emph{recursive}};
        \node at (27,5) {\emph{recursive}};
        \node at (40,5) {\sql{WITH RECURSIVE}};
      \end{scope}
    \end{pgfonlayer}
  \end{tikzpicture}
  \caption{Intermediate forms on the way from~\sql{f} to $\sql{Q}\ssa{f}$.}
  \label{fig:forms}
\end{figure}

\section{Compiling \PLSQL{} Away}
\label{sec:compiling-plsql-away}

The following structures the compilation into a series of transformation
steps. We will use the \PLSQL{} function~\sql{walk}
of~\cref{fig:walk-plsql} as a running example and show interim results
after each step. These intermediate forms of \sql{walk} reveal further
optimizations and simplifications we could apply underway. The four forms
are (also see~\cref{fig:forms}):
\begin{compactdesc}
\item[\step{SSA}] Turn \PLSQL{} function~\sql{f} into \emph{static single assignment}
  (SSA) form.  This maps the diversity of \PLSQL{} control flow constructs to the
  single~\SSA{goto} primitive.
\item[\step{ANF}] From the SSA form derive a functional \emph{administrative normal form} (ANF)
  for~\sql{f} which expresses iteration in terms of (mutually tail-)recursive functions.
\item[\step{UDF}] Flatten mutual recursion and map the ANF functions into one
  tail-recursive \SQL{} user-defined function.
\item[\step{\SQL}] Identify recursive calls and base cases in the body of this UDF
  and embed the body into a template query based on~\sql{WITH~RECURSIVE}. This yields
  the \SQL{} query $\sql{Q}\ssa{f}$ we are after.
\end{compactdesc}
Query $\sql{Q}\ssa{f}$ may then be inlined into~$\sql{Q}$ at the call sites of the
original function~\sql{f}.

\subsection*{\step{\large SSA} Explicit Data Flow and \\
\phantom{\step{\large SSA}} Simple {\Large\SSA{goto}}-Based Control Flow}

Lowering the \PLSQL{} input into its \emph{static single assignment}
(SSA) form~\cite{ssa} preserves the function body's imperative style but
introduces the invariant that any variable is now assigned exactly once
(see~\cref{fig:walk-ssa}). Variable reassignment in the original
function leads to the introduction of a new variable version
(\emph{e.g.}, \sql{step}\ssa{2} in~\cref{line:reassign-step2}) in SSA
form. $\phi$~functions model that an assignment might be reached via
different control flow paths.  The SSA invariant facilitates a wide
range of code simplifications, among these the tracking of redundant
code, constant propagation, or strength reduction.  Others have studied
these in depth~\cite{ssa-opts}. Let us note that \PLSQL{} code is subject to
the same optimizations as any imperative programming language.

Statements in SSA programs are deliberatly simple, featuring
assignments, conditionals,
\SSA{goto}s, and~\SSA{return} only.
In the \PLSQL{} case, expressions in these SSA programs are regular
\SQL{} expressions.  The SSA program contains the original
\sql{walk}'s embedded queries $\sql{Q}\ssa{1\dots 3}$, with their query
parameters instantiated by the appropriate SSA variables
(see~$\sql{Q}\ssa{1}[\sql{location}\ssa{1}]$
in~\cref{line:embed-Q1}, for example).

%% SSA form of walk() PL/SQL function
\begin{figure}[t]
  \centering\small
  \begin{lstbox}
  \begin{lstlisting}[language=ssaanf,escapechar=\%,mathescape=true]
function walk(origin, win, loose, steps)
{
  L0: goto L1;
  L1: reward$\ssa{1}$   $\assign$ %\smash{$\phi(\sql{L0}{:}\,\sql{0}, \sql{L2}{:}\,\sql{reward}\ssa{2})$}%;
      location$\ssa{1}$ $\assign$ %\smash{$\phi(\sql{L0}{:}\,\sql{origin}, \sql{L2}{:}\,\sql{location}\ssa{2})$}%;
      movement$\ssa{1}$ $\assign$ %\smash{$\phi(\sql{L0}{:}\,\sql{'{}'}, \sql{L2}{:}\,\sql{movement}\ssa{2})$}%;
      step$\ssa{1}$     $\assign$ %\smash{$\phi(\sql{L0}{:}\,\sql{0}, \sql{L2}{:}\,\sql{step}\ssa{2})$}%;

      if step$\ssa{1}$ <= steps then                     %\label{line:FOR-start}%
        goto L2
      else return 0;                                     %\label{line:FOR-end}%

  L2: movement$\ssa{2}$ $\assign$ (%\smash{$\mathtt{Q_1}[\sql{location}\ssa{1}]$}%);   %\label{line:embed-Q1}%
      roll $\assign$ random();
      location$\ssa{2}$ $\assign$ (%\smash{$\mathtt{Q_2}[\sql{location}\ssa{1},\sql{movement}\ssa{2},\sql{roll}]$}%);
      reward$\ssa{2}$ $\assign$ (%\smash{$\mathtt{Q_3}[\sql{location}\ssa{2}]$}%);
      step$\ssa{2}$ $\assign$ step$\ssa{1}$ + 1;         %\label{line:reassign-step2}%

      if reward$\ssa{2}$ >= win OR reward$\ssa{2}$ <= loose then
        return step$\ssa{1}$ * sign(reward$\ssa{2}$);
      goto L1;                                           %\label{line:goto-L1}%
}
\end{lstlisting}
  \end{lstbox}
  \caption{Iterative SSA form of \PLSQL{} function~\sql{walk}.}
  \label{fig:walk-ssa}
\end{figure}

Importantly, the zoo of \PLSQL{} control flow
constructs---including~\sql{LOOP},
\sql{EXIT} (to label), \sql{CONTINUE} (at label), \sql{FOREACH}, \sql{FOR}, \sql{WHILE}---
are now exclusively expressed in terms of~\SSA{goto} and jump
labels~\sql{L$x$}. While verbose (the original~\sql{FOR} loop is now
implemented by the conditional~\SSA{goto}
in~\crefrange{line:FOR-start}{line:FOR-end}, the assignment
of~\cref{line:reassign-step2}, and the~\sql{\SSA{goto}~L1}
of~\cref{line:goto-L1}, for example), this homogeneity aids subsequent
steps that trade control flow for function calls.

\subsection*{\step{\large\bfseries ANF} Turning Iteration Into Tail Recursion}

Despite its imperative appearance, the single-assignment restriction
renders SSA already quite close to the functional \emph{administrative
normal form} (ANF)~\cite{ssa-is-fun}. To translate from SSA to ANF
we adapt an algorithm by Chakravarty and colleagues~\cite{ssa-to-anf}.
The resulting programs are purely ex\-press\-ion-based and are composed of---besides basic
subexpressions which, as for SSA, directly follow \SQL{} syntax and
semantics---\ANF{let}(\ANF{rec})$\cdot$\ANF{in} and
\ANF{if}$\cdot$\ANF{then}$\cdot$\ANF{else} expressions only.

Put briefly, we arrive at the ANF of function~\sql{walk} (shown in~\cref{fig:walk-anf})
by
\begin{compactitem}
\item translating each jump label~\sql{L$x$} and the statement block it
  governs into a separate function~\sql{L$x$()},
\item turning~\SSA{goto}~\sql{L$x$} into calls to function~\sql{L$x$()}, while
\item supplying the values of $\phi$-bound variables
  in~\sql{L$x$()} as parameters to these function calls (we additionally
  perform lambda lifting and supply free variables as explicit
  parameters).
\end{compactitem}
If we follow this strategy, iteration (\emph{i.e.}, looping back to a
label) will turn into recursion.  Any such recursive call will be in
\emph{tail position} (since control does not return after a~\SSA{goto};
see the calls to~\sql{L1()} in~\cref{line:call1-L1,line:call2-L1}
and~\sql{L2()} in~\cref{line:call-L2}) which will be crucial in the
final translation to \SQL's~\sql{WITH RECURSIVE}.

Finally, note how sequences of statements have turned into chains of
nested \ANF{let}s which nicely prepares the transcription to a~\SQL{}
UDF in the upcoming step.

%% ANF form of walk() PL/SQL function
\begin{figure}[t]
  \centering\small
  %% measure the width of xx in a listing
  \settowidth{\xx}{\lstinline[columns=fixed]{xx}}\setlength{\x}{0.5\xx}
  \setlength{\y}{\baselineskip}\addtolength{\y}{-1.01pt}
  \begin{tikzpicture}[x=\x,y=-\y,inner sep=0mm]
    %% check width/height of character boxes
    %% \draw[help lines,xstep=1,ystep=-1] (0,0) grid (67,43);
    \node[anchor=north west] at (0,0) {%
      \begin{lstlisting}[language=ssaanf,escapechar=\%,mathescape=true]
function walk(origin, win, loose, steps) =
  letrec L1(reward$\ssa{1}$, location$\ssa{1}$, movement$\ssa{1}$, step$\ssa{1}$) =
    letrec L2(reward$\ssa{1}$, location$\ssa{1}$, movement$\ssa{1}$, step$\ssa{1}$) =
      let movement$\ssa{2}$ = (%\smash{$\mathtt{Q_1}[\sql{location}\ssa{1}]$}%)
      in
        let roll = random()
        in
          let location$\ssa{2}$ = (%\smash{$\mathtt{Q_2}[\sql{location}\ssa{1},\sql{movement}\ssa{2},\sql{roll}]$}%)
          in
            let reward$\ssa{2}$ = reward$\ssa{1}$ + %\smash{$\mathtt{Q_3}[\sql{location}\ssa{2}]$}%
            in
              let step$\ssa{2}$ = step$\ssa{1}$ + 1
              in
                if reward$\ssa{2}$ >= win OR reward$\ssa{2}$ <= loose then
                  step$\ssa{1}$ * sign(reward$\ssa{2}$)
                else
                  L1(reward$\ssa{2}$,location$\ssa{2}$,movement$\ssa{2}$,step$\ssa{2}$)  %\label{line:call1-L1}%
    in
      if step$\ssa{1}$ <= steps then
        L2(reward$\ssa{1}$,location$\ssa{1}$,movement$\ssa{1}$,step$\ssa{1}$)  %\label{line:call-L2}%
      else 0
  in
    L1(0, origin, '', 0)  %\label{line:call2-L1}%
\end{lstlisting}};
    %% visualize letrec scope
    \begin{pgfonlayer}{background}
      \begin{scope}[black!30,line cap=round,line width=0.5pt,xshift=0.5\x,yshift=0.1\y]
        \draw (4,3) -- (4,17);
        \draw (2,2) -- (2,21);
      \end{scope}
    \end{pgfonlayer}
  \end{tikzpicture}
  \caption{Tail-recursive ANF variant of function~\sql{walk}.}
  \label{fig:walk-anf}
\end{figure}

\subsection*{\step{\large\bfseries UDF} Direct Tail Recursion in a \SQL{} UDF}

We take a first step towards \SQL{} and transcribe the intermediate ANF
into a user-defined \SQL{} function (UDF). See~\cref{fig:walk-anfsql}
(ignore the \syn{}~annotations for now). The mutual
recursion between functions~\sql{L1()} and~\sql{L2()} is flattened using
an additional parameter~\sql{fn} whose value discerns between the two
call targets.  This conversion into direct recursion follows standard
defunctionalization tactics~\cite{defunctionalization,funs-are-data-too},
but inlining would work as well.

We follow~\emph{Froid} and compile ANF constructs \ANF{let}$\cdot$\ANF{in} and
\ANF{if}$\cdot$\ANF{then}$\cdot$\ANF{else} into \SQL's table-less~\sql{SELECT}
and~\sql{CASE$\cdot$WHEN}, respectively.  Nested~\ANF{let} bindings translate into
\sql{SELECT}s that are chained using~\sql{LEFT JOIN LATERAL}.  If $\compile{e}$
denotes the \SQL{} equivalent of ANF expression~$e$, we have
$$
\begin{array}{@{}l@{}}
\compile{\ANF{let}\,v\,\sql{=}\,e_1\,\ANF{in}\,e_2} = \\
\qquad \sql{SELECT}~\compile{e_1}~\sql{AS \_(}v\sql{)}~\sql{LEFT~JOIN~LATERAL}~\compile{e_2}~\sql{ON~true} \enspace.
\end{array}
$$
\sql{LATERAL}, introduced by the \SQL:1999 standard, implements the
dependency of~$e_2$ on variable~$v$. In a sense, \sql{LATERAL} thus
assumes the role of statement sequencing via~\sql{;} in \PLSQL.  Here,
\emph{Froid} relied on the Microsoft \SQL{} Server-specific~\sql{OUTER
APPLY} instead~\cite{froid,outer-apply}.  The resulting~\sql{LATERAL}
chains may look intimidating but note that these joins process
\emph{single-row tables} containing bindings of names to (scalar)
values.

This translation step emits a regular \SQL{} UDF which features direct tail
recursion---in the case of~$\sql{walk}\wrapped$ of~\cref{fig:walk-anfsql}
we find two recursive call sites
at~\cref{line:walk1-call-site1,line:walk1-call-site2}.  DBMSs that admit
such recursive UDFs could, in principle, evaluate this function
to compute the result of the original \PLSQL{} procedure.  We observe,
however, that
\begin{compactitem}
\item some DBMSs---among these My\SQL, for example---forbid recursion
  in user-defined \SQL{} functions, and that
\item the direct evaluation of these UDF has disappointing performance
  characteristics.  This is, again, due to significant \Qtof{} and
  \ensuremath{\sql{f}\ttrightarrow\sql{Q}} overhead: the plan for UDF~$\sql{f}\wrapped$'s
  body needs to be prepared and instantiated anew on each recursive invocation.
  Additionally, we quickly hit default stack depth limits, \emph{e.g.}, in
  \Pg{} or \SQL{} Server.
\end{compactitem}

%% SQL formulation of ANF form of walk()
\begin{figure}
  \centering\small
  %% measure the width of xx in a listing
  \settowidth{\xx}{\lstinline[columns=fixed]{xx}}\setlength{\x}{0.5\xx}
  \setlength{\y}{\baselineskip}\addtolength{\y}{-1.01pt}
  \begin{tikzpicture}[x=\x,y=-\y,inner sep=0mm]
    %% check width/height of character boxes
    %% \draw[help lines,xstep=1,ystep=-1] (0,0) grid (67,43);
    \node[anchor=north west] at (0,0) {%
      \begin{lstlisting}[language=sql,escapechar=\%,mathescape=true]
CREATE FUNCTION walk(origin coord, win int, loose int, steps int)
RETURNS int AS %\$\$%
  SELECT walk$\wrapped$(L1, 0, origin, '', 0, win, loose, steps);
%\$\$% LANGUAGE SQL;

CREATE FUNCTION walk$\wrapped$(
  fn int, reward$\ssa{1}$ int, location$\ssa{1}$ coord, movement$\ssa{1}$ text, step$\ssa{1}$ int,
          win int, loose int, steps int)
RETURNS int AS %\$\$%
  SELECT                   %\label{line:body-walk1}\setcounter{listingsline}{\value{lstnumber}}%
    CASE
      WHEN fn = L1 THEN
           CASE
             WHEN step$\ssa{1}$ <= steps THEN
                  walk$\wrapped$(L2,reward$\ssa{1}$,location$\ssa{1}$,movement$\ssa{1}$,step$\ssa{1}$,     %\label{line:walk1-call-site1}%
                           win,loose,steps)
             ELSE 0
           END
      WHEN fn = L2 THEN
           (SELECT
              CASE
                WHEN reward$\ssa{2}$ >= win OR reward$\ssa{2}$ <= loose THEN
                     step$\ssa{1}$ * sign(reward$\ssa{2}$)
                ELSE walk$\wrapped$(L1,reward$\ssa{2}$,location$\ssa{2}$,movement$\ssa{2}$,step$\ssa{2}$,  %\label{line:walk1-call-site2}%
                              win,loose,steps)
              END
            FROM
             (SELECT (%\smash{$\mathtt{Q_1}[\sql{location}\ssa{1}]$}%)) AS _0(movement$\ssa{2}$)
               LEFT JOIN LATERAL
             (SELECT random()) AS _1(roll)
               ON true LEFT JOIN LATERAL
             (SELECT (%\smash{$\mathtt{Q_2}[\sql{location}\ssa{1},\sql{movement}\ssa{2},\sql{roll}]$}%)) AS _2(location$\ssa{2}$)
               ON true LEFT JOIN LATERAL
             (SELECT reward$\ssa{1}$ + (%\smash{$\mathtt{Q_3}[\sql{location}\ssa{2}]$}%)) AS _3(reward$\ssa{2}$)
               ON true LEFT JOIN LATERAL
             (SELECT step$\ssa{1}$ + 1) AS _4(step$\ssa{2}$)
               ON true)
    END
%\$\$% LANGUAGE SQL;
\end{lstlisting}};
    \begin{scope}[xshift=-0.3\x,yshift=0.1\y,cp,color=cp]
      %% simplified computation paths (identify base cases/rec calls at leaves of expression tree)
      \node at ( 2, 9) (toplevel) {\syn{a}};
      \node at ( 4,10) (case1)    {\syn{b}};
      \node at ( 6,11) (when1)    {\syn{c}};
      \node at (11,12) (case2)    {\syn{d}};
      \node at (13,13) (when2)    {\syn{e}};
      \node at (18,14) (walk1)    {\syn{f}};
      \node at (13,16) (else2)    {\syn{g}};
      \node at (18,16) (zero)     {\syn{h}};
      \node at ( 6,18) (when3)    {\syn{i}};
      \node at (11,19) (select)   {\syn{j}};
      \node at (14,20) (case3)    {\syn{k}};
      \node at (16,21) (when4)    {\syn{l}};
      \node at (21,22) (step1)    {\syn{m}};
      \node at (16,23) (else4)    {\syn{n}};
      \node at (21,23) (walk2)    {\syn{o}};
      %% paths
      \draw (toplevel) |- (case1);
      \draw (case1)    |- (when1);
      \draw (when1)    -| (case2);
      \draw (case2)    |- (when2);
      \draw (when2)    -| (walk1);
      \draw (when2)    -- (else2);
      \draw (else2)    -- (zero);
      \draw (when1)    -- (when3);
      \draw (when3)    -| (select);
      \draw (select)   |- (case3);
      \draw (case3)    |- (when4);
      \draw (when4)    -| (step1);
      \draw (when4)    -- (else4);
      \draw (else4)    -- (walk2);
    \end{scope}
  \end{tikzpicture}
  \newsavebox{\paths}\sbox{\paths}{%
    \begin{tikzpicture}[x=5mm,inner sep=0mm,baseline=(syn1.base)]
      \node at (0,0) (syn1) {\syn{}};
      \node at (1,0) (syn2) {\syn{}};
      \draw[cp,color=cp] (syn1) -- (syn2);
    \end{tikzpicture}}
  \newsavebox{\stepSQL}\sbox{\stepSQL}{\step{\SQL}}
  \caption{Recursive \SQL{} UDF~$\sql{walk}\wrapped$ and its wrapper~\sql{walk}. The overlaid
    AST~(\usebox{\paths}) becomes relevant in step~\usebox{\stepSQL}.}
  \label{fig:walk-anfsql}
\end{figure}

\subsection*{\step{\large\bfseries \SQL} Inlinable \SQL{} CTE (\sql{\Large WITH RECURSIVE})}

Instead, we bank on a \SQL:1999-style \emph{recursive CTE}~\cite[\S\,7.12]{sql-1999} as an
evaluation strategy for recursion, ultimately compiling any use
of \PLSQL{} or \SQL{} user-defined functions away. The CTE constructs a
table~\sql{run(call?,$\listof{\sql{args}}$,result)}\footnote{$\listof{\sql{args}}$ abbreviates
the list of UDF arguments. For~\sql{walk}\wrapped, $\listof{\sql{args}} = \sql{fn},
\sql{reward}\ssa{1}, \sql{location}\ssa{1}, \sql{movement}\ssa{1}, \sql{step}\ssa{1},
\sql{win}, \sql{loose}, \sql{steps}$.} that tracks
the evaluation of the recursive UDF~$\sql{f}\wrapped$:

\smallskip\noindent
\begin{tabular}{@{\textbullet\hspace\itemsep}lp{0.82\linewidth}@{}}
  \sql{call?}           & Does the UDF perform a recursive call (\sql{true}) or evaluate a base case (\sql{false})? \\
  $\listof{\sql{args}}$ & In case of a call, what arguments are passed to~$\sql{f}\wrapped$? \\
  \sql{result}          & In a base case, what is the function's result?
\end{tabular}

%% WITH RECURSIVE template for tail recursion
\begin{figure}
  \centering\small
  \begin{lstlisting}[language=sql,escapechar=\%,mathescape=true]
  WITH RECURSIVE run("call?", $\listof{\sql{args}}$, result) AS (
    -- original function call
    SELECT true AS "call?", $\listof{\sql{\vrule height 1.4mm width 0mm\smash{in}}}$ AS $\listof{\sql{args}}$, CAST(NULL AS $\tau$) AS result  %\label{line:start-evaluation}%
      UNION ALL
    -- subsequent recursive calls and base cases
    SELECT iter.*
    FROM   run AS r,
           LATERAL ($\var{body}(\sql{f}\wrapped,\sql{r})$) AS iter("call?", $\listof{\sql{args}}$, result)  %\label{line:body-evaluation}%
    WHERE  r."call?"  %\label{line:continue-evaluation}%
  )
  -- extract result of final recursive function invocation
  SELECT r.result        %\label{line:stop-evaluation1}%
  FROM   run AS r
  WHERE  NOT r."call?"   %\label{line:stop-evaluation2}%
\end{lstlisting}
  \caption{\SQL{} CTE template that evaluates tail-recursive~$\sql{f}\wrapped$.}
  \label{fig:tail-rec-template}
\end{figure}

\smallskip\noindent
Recall that we are dealing with tail recursion: once we reach a base case,
the UDF's result is known and no further recursive ascent is required. The
obtained result may thus be returned as the original function's outcome.

\smallskip\noindent
The evaluation of call $\sql{f(\listof{\sql{\vrule height 1.8mm width 0mm\smash{in}}})}$
is expressed by the simple~\sql{WITH RECURSIVE}
\SQL{} code template of~\cref{fig:tail-rec-template}:
\begin{compactitem}
\item \cref{line:start-evaluation}: Start evaluation with the original invocation of the
  UDF for argument list~$\listof{\sql{\vrule height 1.8mm width 0mm\smash{in}}}$. $\sql{f}\wrapped$'s result (of type~$\tau$) is yet unknown
  and thus encoded as~\sql{NULL}.
\item \cref{line:continue-evaluation}: Continue evaluation as long as new recursive calls are to
  be performed.
\item \cref{line:body-evaluation}:  Evaluate the body of~$\sql{f}\wrapped$ for the current arguments held
  in~\sql{r.$\listof{\sql{args}}$}. This either leads to a new call or the evaluation of a base case.
\item \crefrange{line:stop-evaluation1}{line:stop-evaluation2}: Once we reach a base, extract its
  result. Done.
\end{compactitem}

\smallskip\noindent
The code template of~\cref{fig:tail-rec-template} is entirely generic.
It is to be completed with a slightly adapted body---$\var{body}(\sql{f}\wrapped,\sql{r})$---of
  the UDF~$\sql{f}\wrapped$ for~\sql{f}. In this adaptation,
\begin{compactitem}
\item a recursive call~$\sql{f\wrapped($\listof{\var{args}}$)}$ is replaced by the construction of
  row $\sql{(true,$\listof{\var{args}}$,NULL)}$ which encodes just that call in simulation table~\sql{run},
\item a base case expression with result~$v$ of type~$\tau$ is replaced by row
  $\sql{(false,NULL,$v$)}$.
\end{compactitem}

\medskip\noindent
\cref{fig:walk-body} depicts the resulting
body~$\var{body}(\sql{walk}\wrapped,\sql{r})$ for the recursive UDF
of~\cref{fig:walk-anfsql}.  The construction
of~$\var{body}(\sql{f}\wrapped,\cdot)$ calls
for a simple abstract syntax tree (AST) traversal of the body of UDF~$\sql{f}\wrapped$. Selected
fragments of the AST for function~$\sql{walk}\wrapped$ are shown in an overlay
of~\cref{fig:walk-anfsql}.  This traversal identifies the leaves of the
computation---\emph{i.e.}, the recursive call sites \syn{f}, \syn{o} and
base case expressions~\syn{h}, \syn{m}---and performs the local
replacements described above.

%% translated body of SQL UDF formulation of ANF
\begin{figure}
  \centering\small
  %% measure the width of xx in a listing
  \settowidth{\xx}{\lstinline[columns=fixed]{xx}}\setlength{\x}{0.5\xx}
  \setlength{\y}{\baselineskip}\addtolength{\y}{-1.01pt}
  \begin{tikzpicture}[x=\x,y=-\y,inner sep=0mm]
    %% check width/height of character boxes
    %% \draw[help lines,xstep=1,ystep=-1] (0,0) grid (67,43);
    \node[anchor=north west] at (0,0) {%
  \begin{lstlisting}[language=sql,escapechar=\%,mathescape=true,firstnumber=\thelistingsline]
SELECT
  CASE
    WHEN r.fn = L1 THEN
         CASE
           WHEN r.step$\ssa{1}$ <= r.steps THEN
                ROW(true,                        %\suppressnumber%
                    ROW(L2,r.reward$\ssa{1}$,r.location$\ssa{1}$,r.movement$\ssa{1}$,r.step$\ssa{1}$,
%\raisebox{3pt}{\gutterlineskip}%                           r.win, r.loose, r.steps),
                    NULL)                        %\addtocounter{lstnumber}{1}\reactivatenumber%
           ELSE ROW(false, NULL, 0)
         END
    WHEN r.fn = L2 THEN
         (SELECT
            CASE
              WHEN reward$\ssa{2}$ >= r.win OR reward$\ssa{2}$ <= r.loose THEN
                   ROW(false,NULL,r.step$\ssa{1}$ * sign(reward$\ssa{2}$))
              ELSE ROW(true,                     %\suppressnumber%
                       ROW(L1,reward$\ssa{2}$,location$\ssa{2}$,movement$\ssa{2}$,step$\ssa{2}$,
%\raisebox{3pt}{\gutterlineskip}%                              r.win, r.loose, r.steps),
                       NULL)                     %\addtocounter{lstnumber}{1}\reactivatenumber%
            END
          FROM
           (   -- code of %\lst@commentstyle\cref{fig:walk-anfsql}% %\suppressnumber%
%\raisebox{3pt}{\gutterlineskip}%            %\smash{\raisebox{2pt}{$\vdots$}}%) -- %\lst@commentstyle (binds reward\ssa{2},location\ssa{2},movement\ssa{2},step\ssa{2})% %\addtocounter{lstnumber}{9}\reactivatenumber%
  END
\end{lstlisting}};
    \begin{scope}[xshift=-0.3\x,yshift=0.1\y,cp,color=cp]
      %% encoding of base cases/rec calls at leaves of expression tree
      \node at (16, 5) (walk1)    {\syn{f}};
      \node at (16, 9) (zero)     {\syn{h}};
      \node at (19,15) (step1)    {\syn{m}};
      \node at (19,16) (walk2)    {\syn{o}};
    \end{scope}
  \end{tikzpicture}
  \caption{Adapted UDF body~$\var{body}(\sql{walk}\wrapped, \sql{r})$.
    At \syn{f}, \syn{o}, and \syn{h}, \syn{m} row construction replaces recursive calls and base cases.}
  \label{fig:walk-body}
\end{figure}

\medskip\noindent
\textbf{Finalization.} A merge of~$\var{body}(\sql{f}\wrapped,\sql{r})$
with the \SQL{} code template yields a pure \SQL{} expression which may
be inlined at~\sql{f}'s call sites in the embracing query~\sql{Q}. Any
occurrence of \PLSQL{} has been compiled away. The DBMS will be able to
compile the resulting \SQL{} query into a regular plan and jointly
optimize the formerly separated code of~\sql{Q}, the transformed body
of~\sql{f}, and the embedded queries~$\sql{Q}_i$. Most importantly, the
evaluation of~\sql{Q} instantiates this joint plan \emph{once} and will
proceed without the need for~\Qtof{} or~\ftoQ{} context switches.  The
upcoming section quantifies the benefits we can now reap.

\smallskip\noindent
\textbf{Beyond tail recursion.}
Let us close this discussion by noting that the
\sql{WITH RECURSIVE}-based simulation of a recursive function
extends beyond tail recursion.  Table~\sql{run} can be generalized to
hold a true \emph{call graph} that then does support recursive ascent. While
this is not needed in the context of the present work, this paves the
way for an intuitive, functional-style notation for \SQL{} UDFs that may
employ linear or general $n$-way recursion.  The run time savings
can be---again, due to the absence of plan instantiation effort---significant.
We are actively pursuing this idea in a parallel strand of research
that aims to leave more complex recursive compuation in the hands of the DBMS itself.

\section{Once \PLSQL{} is Gone}
\label{sec:experiments}

Function \sql{walk()} is not the exception.  The described overheads are
pervasive~\cite{froid} and we, too, observed them across a variety of
\PLSQL{} functions.

%% PL/SQL overheads for various example functions
\begin{table}
  \centering\small
  \caption{Run time spent (in\,\%) during \PLSQL{} evaluation. Bold entries
    indicate context switch overhead of kind~\ftoQ.}
  \begin{narrow}{-1mm}{-1mm}
    \begin{tabular}{r>{\bfseries}c.>{\bfseries}c.}
      \toprule
      \textbf{\PLpgSQL{} Function} &
      \multicolumn{1}{r}{\sql{Exec$\cdot$Start}} &
      \multicolumn{1}{r}{\sql{Exec$\cdot$Run}} &
      \multicolumn{1}{r}{\sql{Exec$\cdot$End}} &
      \multicolumn{1}{r}{\textbf{Interp}} \\
      \cmidrule(lr){1-1}\cmidrule(lr){2-5}
      \sql{walk}      & 30.89                      & 55.13 & 4.36           &  9.63 \\[-1.5mm]
      \tiny see~\cref{fig:walk-plsql} \\
      \sql{parse}     & 13.84                      & 68.52 & 2.20           & 15.62 \\[-1.5mm]
      \tiny via finite state automaton \\
      \sql{traverse}  & 31.80                      & 35.82 & 6.03           & 26.35 \\[-1.5mm]
      \tiny directed graph traversal \\
      \sql{fibonacci} &  \phantom{0}0\phantom{.00} & 90.45 & 0\phantom{.00} &  9.55 \\[-1.5mm]
      \tiny iteratively compute $\var{fib}(n)$ \\
      \bottomrule
    \end{tabular}
  \end{narrow}
  \label{tab:overheads}
\end{table}

\smallskip\noindent
\textbf{Context switching overhead.}
\cref{tab:overheads} contains a sample of iterative functions
and reports the run time for repeated plan instantiation and
deallocation to evaluate their embedded queries.
Columns~\sql{Exec$\cdot$Start} and~\sql{Exec$\cdot$End} embody the
\ftoQ{} context switch overhead present in \Pg{}
(recall~\cref{sec:introduction}). Across the functions, we find overall
\ftoQ{} overheads of up to~$38\%$.  Only the
columns~\sql{Exec$\cdot$Run} and~\textbf{Interp} represent productive
evaluation effort: the execution of embedded queries and \PLSQL{}
interpretation, respectively.  Function~\sql{fibonacci}, an iterative
computation of the $n$th Fibonacci number, evaluates arithmetic
expressions only and does not execute embedded queries.
\Pg{} evaluates such \emph{simple expressions} using a fast path that
already foregoes plan instantiation.  Compiling
\PLSQL{} away does not promise much in this case.  Still, turning
query-less iterative functions into pure \SQL{} can uncover
opportunities for parallel evaluation---this is a direction we
have not yet explored.

\begin{figure}
  \centering\small
  \newsavebox{\stopwatch}\sbox{\stopwatch}{\StopWatchStart}
  \begin{tikzpicture}[x=5mm,y=0.008mm]
    %% axes
    \foreach \y/\t in {0/0,500/,1000/1000,1500/,2000/2000,2500/,3000/3000,3500/,4000/4000}
      { \draw[black!50,densely dotted] (-0.3,\y) -- (11,\y);
        \node[left] at (-0.3,\y) {\ifx\t\empty\else\pgfmathprintnumber[1000 sep=\,]{\t}\fi};
      }
    \node at (-1,5000) {\usebox{\stopwatch} \raisebox{2pt}{[ms]}};
    \foreach \x/\i in {1/10,2/,3/,4/,5/50,6/,7/,8/,9/,10/100}
      { \draw (\x,0) -- (\x,-200);
        \node at (\x,-500) {\ifx\i\empty\else\pgfmathprintnumber[1000 sep=\,]{\i}\fi};
      }
    \node at (5,-1000) {\#\,iterations ($\times 1\,000$)};
    \begin{scope}[thick,->]
      \draw (0,0) -- (11,0);
      \draw (0,0) -- ( 0,4500);
    \end{scope}
    %% measurements
    %% min/max envelopes
    \begin{scope}[envelope]
      %% WITH REC
      \fill (1,225.9) foreach \x/\rec in {2/441.4,3/668.4,4/879.6,5/1054.8,6/1321.7,7/1436.4,8/1760.2,9/1951.6,10/2160.4,
                                          10/2286.7,9/2055.1,8/1816.5,7/1594.3,6/1369.0,5/1145.0,4/930.0,3/716.6,2/459.4,1/257.9}
        { -- (\x,\rec) } -- cycle;
      %% PL/SQL
      \fill (1,397.7) foreach \x/\pl  in {2/799.0,3/1195.9,4/1592.1,5/1962.7,6/2382.8,7/2687.9,8/3159.5,9/3501.7,10/3885.9,
                                          10/4021.9,9/3609.1,8/3201.3,7/2800.5,6/2411.1,5/2036.6,4/1638.5,3/1242.3,2/811.6,1/412.2}
      { -- (\x,\pl) } -- cycle;
    \end{scope}
    %% averages
    %% PL/SQL
    \foreach \pl [count=\x from 1] in {406.7,804.5,1218.9,1604.9,2001.2,2394.1,2729.4,3178.9,3567.5,3975.6}
      { \draw[fill=white] (\x,\pl) circle (1pt);
      }
    %% WITH REC
    \foreach \rec [count=\x from 1] in {228.1,455.1,689.4,908.1,1123.1,1356.8,1534.7,1806.6,2009.0,2218.3}
      { \fill[draw] (\x,\rec) circle (1pt);
      }
    %% savings
    \begin{scope}[font=\scriptsize,black!60]
    \draw[shorten >=1,shorten <=1,<-] (8,1806.6) -- (8,3178.9)
          node[right,pos=0.7] {$43\%$};
    \end{scope}
    %% legend
    \begin{scope}[font=\scriptsize]
      \draw[fill=white] (1,4250) circle (1pt) node[right=1mm] {\PLSQL};
      \fill[draw]       (1,3750) circle (1pt) node[right=1mm] {\sql{WITH RECURSIVE}};
      %% min/max envelope
      \begin{scope}[envelope]
        \fill (0.7,3160) -- (0.9,3220) -- (1.2,3230) -- (1.2,3340) -- (0.9,3260) -- (0.7,3270) -- cycle;
      \end{scope}
      \node[right=1mm] at (1,3250) {min/max envelope};
    \end{scope}
  \end{tikzpicture}
  \caption{Iterative vs. recursive: wall clock time for~\sql{walk()} on \Pg~11.3
    across varying intra-function iterations.}
  \label{fig:performance-walk}
\end{figure}
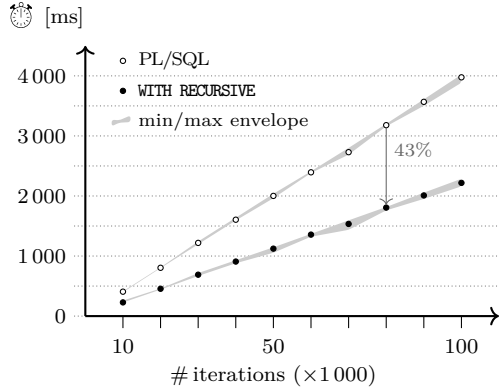

\smallskip\noindent
\textbf{Iterative \PLSQL{} vs.\ Recursive \SQL.}
For \PLpgSQL{} function~\sql{walk()}, \cref{tab:overheads} indicates
potential run time savings of about $35\% \approx 30.89\% + 4.36\%$ should
we manage to get rid of context switching overhead.  The translation from
iterative \PLSQL{} to pure \SQL{} built on a recursive CTE can indeed realize this
advantage. \cref{fig:performance-walk} shows the wall clock time
of one invocation of~\sql{walk()} for a growing number of \sql{FOR}~loop
iterations (which is controlled by parameter~\sql{steps}, see~\cref{fig:walk-plsql}).
Throughout the experiment, the recursive \SQL{} variant consistently
shows an even greater run time savings of approximately~$43\%$. Beyond saved
context switches, this suggests that the evaluation of pure \SQL{} expressions
generally undercuts the interpretation of \PLSQL{} statements.

%% performance of walk() (PL/SQL vs. WITH REC for growing # of iterations)

(The measurements of~\cref{fig:performance-walk} have been taken on a \Pg~11.3 instance hosted on
a Linux-based x86 box with 8 Intel Core\raisebox{1ex}{\tiny\textls[-80]{TM}}\,i7 CPUs
running at $3.66\,\text{GHz}$ with~64\,GB of~RAM.  We report the average
as well as the window of \mbox{minimal/maximal} measurements of ten~runs.)

%% All measurements on PgSQL 11.3, on Linux host Linux 5.1, x86-64, 64 GB RAM, 8x Intel Corei7 @ 3.6GHz, HDD 1.8TB

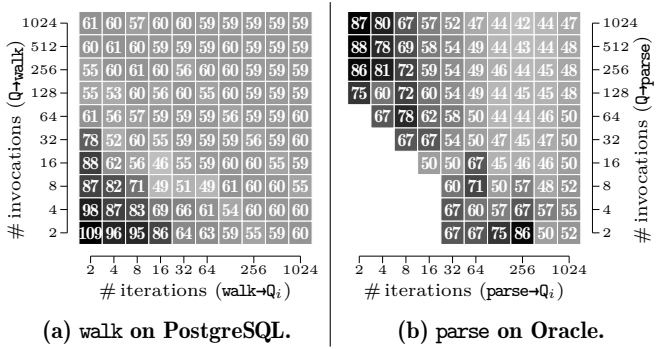
\begin{figure}
  \begin{narrow}{-2mm}{-2mm}
  \centering\small
  %% gray color map based on height on map
  \pgfplotsset{colormap={heat}{
    color(  0)=(black!25)
    color(100)=(black!100)}}
  \begin{subfigure}[b]{0.49\linewidth}
    \begin{tikzpicture}[x=\G,y=\G]
      %% axes
      \begin{scope}[very thin,inner sep=1pt,font=\tiny]
        \begin{scope}[yshift=-\G]
          \draw (0,0) -- (9,0);
          \foreach \x in {1,2,3,4,5,6,8,10}
            { \draw (\x-1,0) -- (\x-1,-0.3)
                 node[below] {\pgfmathparse{scalar(2^\x)}
                              \pgfmathprintnumber[precision=0,1000 sep=]{\pgfmathresult}};
            }
          \node[font=\scriptsize] at (4.5,-1.5) {\#\,iterations (\sql{walk}\raisebox{-1pt}{$\ttrightarrow$}$\sql{Q}_i$)};
          \end{scope}
        \begin{scope}[xshift=-\G]
          \draw (0,0) -- (0,9);
          \foreach \y in {1,2,3,4,5,6,7,8,9,10}
            { \draw (0,\y-1) -- (-0.3,\y-1)
                 node[left] {\pgfmathparse{scalar(2^\y)}
                             \pgfmathprintnumber[precision=0,1000 sep=]{\pgfmathresult}};
            }
          \end{scope}
          \node[rotate=90,font=\scriptsize] at (-3.2,4) {\#\,invocations ($\sql{Q}$\raisebox{-1pt}{$\ttrightarrow$\sql{walk}})};
      \end{scope}
      %% heat map
      \begin{axis}[x=\G,y=\G,at={(0,0)},anchor={origin},hide axis]
        \addplot[%
          scatter,
          only marks,
          scatter src=explicit,
          scatter/use mapped color={fill=mapped color},
          mark=square*,mark size=1/2*\G,mark options={draw=white}]
          table[meta expr={100*\thisrow{rec}/\thisrow{pl}}]          %% heat ≡ PL/SQL to WITH REC ratio
          {assets/heat-map-pg.xyh};
        \addplot[%
          scatter,
          nodes near coords={\textls[-80]{\bfseries\pgfmathprintnumber[assume math mode=true,precision=0]{\pgfplotspointmeta}}},
          nodes near coords align={anchor=center},
          nodes near coords style={font=\smaller,text=white},
          scatter src=explicit,
          mark=none]
          table[meta expr={100*\thisrow{rec}/\thisrow{pl}}]          %% heat ≡ PL/SQL to WITH REC ratio
          {assets/heat-map-pg.xyh};
      \end{axis}
    \end{tikzpicture}%
    \caption{\sql{walk} on \Pg.}
    \label{fig:heat-map-walk}
  \end{subfigure}%
  \vrule\hskip1mm
  \begin{subfigure}[b]{0.49\linewidth}
    \begin{tikzpicture}[x=\G,y=\G]
      %% axes
      \begin{scope}[very thin,inner sep=1pt,font=\tiny]
        \begin{scope}[yshift=-\G]
          \draw (0,0) -- (9,0);
          \foreach \x in {1,2,3,4,5,6,8,10}
            { \draw (\x-1,0) -- (\x-1,-0.3)
                 node[below] {\pgfmathparse{scalar(2^\x)}
                              \pgfmathprintnumber[precision=0,1000 sep=]{\pgfmathresult}};
            }
          \node[font=\scriptsize] at (4.5,-1.5) {\#\,iterations (\sql{parse}\raisebox{-1pt}{$\ttrightarrow$}$\sql{Q}_i$)};
          \end{scope}
        \begin{scope}[xshift=-\G]
          \draw (11,0) -- (11,9);
          \foreach \y in {1,2,3,4,5,6,7,8,9,10}
            { \draw (11,\y-1) -- (11.3,\y-1)
                 node[right] {\pgfmathparse{scalar(2^\y)}
                             \pgfmathprintnumber[precision=0,1000 sep=]{\pgfmathresult}};
            }
          \end{scope}
          \node[rotate=90,font=\scriptsize] at (12.2,4) {\#\,invocations ($\sql{Q}$\raisebox{-1pt}{$\ttrightarrow$\sql{parse}})};
      \end{scope}
      %% heat map
      \begin{axis}[x=\G,y=\G,at={(0,0)},anchor={origin},hide axis]
        \addplot[%
          scatter,
          only marks,
          scatter src=explicit,
          scatter/use mapped color={fill=mapped color},
          mark=square*,mark size=1/2*\G,mark options={draw=white}]
          table[meta expr={100*\thisrow{ratio}}]          %% heat ≡ PL/SQL to WITH REC ratio
          {assets/heat-map-ora.xyh};
        \addplot[%
          scatter,
          nodes near coords={\textls[-80]{\bfseries\pgfmathprintnumber[assume math mode=true,precision=0]{\pgfplotspointmeta}}},
          nodes near coords align={anchor=center},
          nodes near coords style={font=\smaller,text=white},
          scatter src=explicit,
          mark=none]
          table[meta expr={100*\thisrow{ratio}}]          %% heat ≡ PL/SQL to WITH REC ratio
          {assets/heat-map-ora.xyh};
      \end{axis}
    \end{tikzpicture}%
    \caption{\sql{parse} on \Oracle.}
    \label{fig:heat-map-parse}
  \end{subfigure}
  \end{narrow}
  \caption{Relative run time (in~\%) of~recursive \SQL{} vs.\ iterative
    \PLSQL{}. Light colors indicate an advantage for \SQL.}
  \label{fig:heat-maps}
\end{figure}

\smallskip\noindent
\textbf{Scaling the number of context switches.}
We can quite consistently observe these savings of $\geqslant 40\%$
across a wide range of scenarios. \cref{fig:heat-map-walk} varies the
number of invocations of~\sql{walk()} as well as the
intra-function~\sql{FOR}~loop iterations to obtain a heat map of run
time improvements.  Only very small numbers of
invocations and/or iterations fail to compensate the one-time cost to
optimize and evaluate the template query
of~\cref{fig:tail-rec-template} (see the heat map's lower left). Beyond 32~invocations and/or iterations,
the transformation to recursive \SQL{} always is a clear win.

\smallskip\noindent
\textbf{Beyond \Pg.} Modulo syntactic details, we were able to apply the
function transformation of~\cref{sec:compiling-plsql-away} immediately
to Oracle, My\SQL, \SQL{} Server, and HyPer. As an example,
\cref{fig:heat-map-parse} shows how the evaluation of~\sql{parse()} on
\Oracle{} can significantly benefit once
\PLSQL{} is traded for recursive \SQL{} (measurements in the lower left
appeared to be close to\,%
\begin{tikzpicture}[x=\G,y=\G,inner sep=1pt,baseline=(onehundred.base)]
  \fill[black] (0,0) rectangle (1,1);
  \node[font=\smaller,text=white] at (0.5,0.5) (onehundred) {\textls[-100]{\bfseries 100}};
\end{tikzpicture};
we have omitted them here due to the DBMS's coarse timer resolution).
\SQL{}ite3 lacks support for~\sql{LATERAL}, but a simple syntactic rewrite
brought the functions to run on a system that formerly lacked any
support for \PLSQL{} at all.  Compiling \PLSQL{} away could, generally,
pave the way to provide scripting support for more database engines.

%% can observe Q→f and f→Qᵢ across a variety of RDBMS (Pg, Oracle, MySQL, SQL Server)
%% saving context switch overhead => vary # of Q→f, f→Qᵢ transitions to explore behavior of PL/SQL vs. WITH REC
%%   -> heat map
%%   - transformation to WITH REC always a win, stable 60% in high-volume computation (exception: VERY low # of iterations)
%%   - not specific to Pg: see heat map for parse() in Oracle 19c
%%     - empty fields in heat map: granularity of timer not sufficient for few invocations/few iterations
%%       (eval times virtually equal)
%% - can evaluate these functions also on SQLite3 (no PL/SQL) and HyPer (no PL/SQL, but HyperScript)
%%   - comparison with HyPerScript work to be done (→ WITH ITERATE!)

%% heat map (performance improvement for walk() over PL/SQL function)
%% varying #invocations (Q→f) AND #iterations (f→Qᵢ)

\smallskip\noindent
\mbox{\textbf{When\! \sql{WITH\! RECURSIVE} does\! too\! much.\! Exploiting\! tail\! recursion.}} \newline
The transformation from~SSA to~ANF compiles~\SSA{goto} into \emph{tail}
recursion which obviates the need for recursive ascent: any activiation
of a tail-recursive function already contains its complete evaluation
context---typically held in the function's arguments.  Tail recursion, thus,
needs no stack (fra\-mes). Vanilla \sql{WITH
RECURSIVE}, however, collects a trace of all function invocations and
their respective arguments (recall table~\sql{run}
of~\cref{fig:tail-rec-template}). For our purposes, accumulating this
trace is wasted effort and the template of~\cref{fig:tail-rec-template}
indeed uses the predicate
\sql{NOT r."call?"} in~\cref{line:stop-evaluation2} to dispose of the trace
and hold on to the function's \emph{final} activation only.

Here, a hypothetical ``\sql{WITH TAIL RECURSIVE}'' that keeps the most
recent~\sql{run} row only would be a better fit.  Interestingly, earlier
work on the evaluation of complex analytics in HyPer has described just
this construct, coined~\sql{WITH ITERATE} in~\cite{hyper-iterate}.
To assess the benefit in the context of \PLSQL{} elimination, we implemented
\sql{WITH ITERATE} in \Pg~11.3.  The resulting space savings can indeed
be profound, in particular for functions with potentially sizable
arguments. One such example is function~\sql{parse()} which receives its
input text as an argument. \cref{tab:buffer-writes} lists the number of
buffer page writes performed by \Pg{} when inputs of growing length are
parsed. \sql{WITH ITERATE} realizes the promise of tail recursion and
requires no space at all, while \sql{WITH RECURSIVE} exhibits quadratic
space appetite (both, the number of required iterations that consume
one input character each as well as the lengths of the residual strings
left to parse, do grow).

In an age of complex in-database computation, we step forward and propose
that a construct like~\sql{WITH ITERATE} should find its way into the \SQL{} standard.

%% buffer writes during evaluation of parse() with WITH ITERATE/WITH REC
\begin{table}
  \centering\small
  \caption{Eliminating buffering effort via~\sql{WITH ITERATE}.}
  \begin{tabular}{rcr@{~}p{2.5cm}}
    \toprule
    \textbf{\#\,Iterations} & \multicolumn{3}{c}{\textbf{\#\,Buffer Page Writes}} \\
     (= input length)       & \sql{WITH ITERATE} & \multicolumn{2}{c}{\sql{WITH RECURSIVE}} \\
    \cmidrule(lr){1-1}\cmidrule(lr){2-4}
    $10\,000$ & $0$ &   $6\,132$ & \tikz[remember picture,baseline=-2pt] \node[inner sep=0mm] (p10000) {}; \\
    $20\,000$ & $0$ &  $24\,471$ & \tikz[remember picture,baseline=-2pt] \node[inner sep=0mm] (p20000) {}; \\
    $30\,000$ & $0$ &  $55\,016$ & \tikz[remember picture,baseline=-2pt] \node[inner sep=0mm] (p30000) {}; \\
    $40\,000$ & $0$ &  $97\,769$ & \tikz[remember picture,baseline=-2pt] \node[inner sep=0mm] (p40000) {}; \\
    $50\,000$ & $0$ & $152\,729$ & \tikz[remember picture,baseline=-2pt] \node[inner sep=0mm] (p50000) {}; \\
    \bottomrule
  \end{tabular}
  \begin{tikzpicture}[x=1.6mm,remember picture,overlay,inner sep=0mm]
    \begin{scope}[line width=1pt,shorten >=-0.6132mm,-{Circle[length=3pt]}]
      \draw (p10000) -- +( 0.6132,0) node (p1) {};
      \draw (p20000) -- +( 2.4471,0) node (p2) {};
      \draw (p30000) -- +( 5.5016,0) node (p3) {};
      \draw (p40000) -- +( 9.7769,0) node (p4) {};
      \draw (p50000) -- +(15.2729,0) node (p5) {};
    \end{scope}
    \begin{scope}[densely dotted]
      \draw (p1) -- (p2);
      \draw (p2) -- (p3);
      \draw (p3) -- (p4);
      \draw (p4) -- (p5);
    \end{scope}
  \end{tikzpicture}
  \label{tab:buffer-writes}
\end{table}

\section{(Too Early For) Conclusions}
\label{sec:un-conclusions}

This marks the beginning of a thread of research in which we aim to
explore fresh ways to support complex in-database computation,
preferably \emph{without} turning existing engines on their head.
Directions waiting to be explored include:
\begin{compactitem}
\item With its imminent Version~12, \Pg{} will offer hooks that
  enable merging of CTEs with their containing queries.
  Inlining compiled functions into their calling query then opens up
  additional optimization opportunities.
\item Flattening nested iteration into flat recursion facilitates
  efficient evaluation through partitioning and pa\-rallel\-ism.
\item Given their proper encoding in table~\sql{run}, function return
  types and variables that are table-valued (the latter are not allowed
  in~\PLSQL) can be supported by the compilation scheme.
\item Beyond \PLSQL: With its ability to compile arbitrary SSA
  programs, this provides the groundwork required for the
  compilation and evaluation of expressive imperative languages
  within regular DBMSs and thus close to the data.
\end{compactitem}

%% ....................................................................
%% bibliography

\bibliographystyle{abbrv}
\begin{spacing}{0.8}
\bibliography{compile-plsql-away}

\begin{thebibliography}{10}

\bibitem{ssa}
B.~Alpern, M.~Wegman, and F.~Zadeck.
\newblock {Detecting} {Equality} of {Values} in {Programs}.
\newblock In {\em Proc.\ POPL}, 1988.

\bibitem{ssa-is-fun}
A.~Appel.
\newblock {SSA} is {Functional} {Programming}.
\newblock {\em ACM SIGPLAN Notices}, 33(4), 1998.

\bibitem{ssa-to-anf}
M.~Chakravarty, G.~Keller, and P.~Zadarnowski.
\newblock {A} {Functional} {Perspective} on {SSA} {Optimisation} {Algorithms}.
\newblock {\em Electronic Notes in Theoretical Computer Science}, 82(2), 2004.

\bibitem{ssa-opts}
R.~Cytron, J.~Ferrante, B.~Rosen, M.~Wegman, and F.~Zadeck.
\newblock {Efficiently} {Computing} {Static} {Single} {Assignment} {Form} and
  the {Control} {Dependence} {Graph}.
\newblock {\em ACM TOPLAS}, 13(4), 1991.

\bibitem{outer-apply}
C.~Galindo-Legaria and M.~Joshi.
\newblock {Orthogonal} {Optimization} of {Subqueries} and {Aggregation}.
\newblock In {\em Proc.\ SIGMOD}, 2001.

\bibitem{funs-are-data-too}
T.~Grust, N.~Schweinsberg, and A.~Ulrich.
\newblock {Functions} are {Data} {Too} ({Defunctionalization} for {\PLSQL}).
\newblock In {\em Proc.\ VLDB}, 2013.

\bibitem{brent-ozar}
B.~Ozar.
\newblock {Froid:} {How} {\SQL} {Server} 2019 {Might} {Fix} the {Scalar}
  {Functions} {Problem}.
\newblock \url{www.brentozar.com}, 2018.

\bibitem{hyper-iterate}
L.~Passing, M.~Then, N.~Hubig, H.~Lang, M.~Schreier, S.~G{\"u}nnemann,
  A.~Kemper, and T.~Neumann.
\newblock {\SQL-} and {Operator-Centric} {Data} {Analytics} in {Relational}
  {Main-Memory} {Databases}.
\newblock In {\em Proc.\ EDBT}, 2017.

\bibitem{froid}
K.~Ramachandra, K.~Park, K.~Emani, A.~Halverson, C.~Galindo-Legaria, and
  C.~Cunningham.
\newblock {Froid:} {Optimization} of {Imperative} {Programs} in a {Relational}
  {Database}.
\newblock {\em Proc.\ VLDB}, 11(4), 2018.

\bibitem{defunctionalization}
J.~Reynolds.
\newblock {Definitional} {Interpreters} for {Higher-Order} {Programming}
  {Languages}.
\newblock {\em Higher-Order and Symbolic Computation}, 11, 1998.

\bibitem{sql-1999}
{\em {SQL:1999}. {Database} {Languages}--{\SQL}--{Part 2}: {Foundation}}.
\newblock {ISO/IEC}~9075-2:1999.

\end{thebibliography}
\end{spacing}

\end{document}